\documentclass[12 pt, amssymb,prb,showpacs]{revtex4}
\input{psfig.sty}
\usepackage{epsfig}
\usepackage{dcolumn}
\usepackage{amsmath}
\usepackage[a4paper,dvips]{geometry}
\begin{document}
\title{\bf\ Magnetoresistivity Modulated Response in Bichromatic
Microwave Irradiated Two Dimensional Electron Systems}
\author{J. I\~narrea}
\affiliation{Escuela Polit\'ecnica Superior, Universidad Carlos
III, Leganes, Madrid, 28911, Spain\\ Instituto de Ciencia de
Materiales, CSIC, Cantoblanco, Madrid, 28049, Spain}
\author{G. Platero} \affiliation{Instituto de Ciencia de Materiales,
CSIC, Cantoblanco, Madrid, 28049, Spain}
\date{\today}

\begin{abstract}
We analyze the effect of bichromatic microwave irradiation on the
magnetoresistivity of a two dimensional electron system. We follow
the model of microwave driven Larmor orbits in a regime where two
different microwave lights with different frequencies are
illuminating the sample ($w_{1}$ and $w_{2}$). Our calculated
results demonstrate that now the electronic orbit centers are driven
by the superposition of two harmonic oscillatory movements with the
frequencies of the microwave sources. As a result the
magnetoresisitivity response presents modulated pulses in the
amplitude with a frequency  of $\frac{w_{1}-w_{2}}{2}$, whereas the
main response oscillates with $\frac{w_{1}+w_{2}}{2}$.
\end{abstract}

\maketitle

\newpage

In the field of Condensed Matter Physics, very few problems have
produced such a intense activity, experimental and theoretical, like
Microwave Induced Resistivity Oscillations
(MIRO)\cite{zudov,studenikin}  and Zero Resistance States
(ZRS)\cite{mani,zudov2}. From the experimental standpoint,
remarkable contributions are being published in a continuous basis.
Among them we can highlight activated temperature dependence in the
magnetoresisitivity ($\rho_{xx}$)
response\cite{mani,zudov2,willett}, quenching of $\rho_{xx}$
response at high microwave (MW) intensities \cite{mani2,studenikin},
absolute negative conductivity (ANC) and breakdown of ZRS
\cite{willett,mani2,studenikin,zudov3}, suppression of MIRO and ZRS
by in-plane magnetic field \cite{yang,mani3} and the evidence that
MIRO and ZRS are notably immune to the sense of circular
polarization of MW radiation\cite{smet}. Very recently, an
experimental achievement has joined the group of new contributions.
In this case an ultraclean  2DES is subjected to bichromatic MW
radiation coming from two monochromatic sources with different
frequencies $w_{1}$ and $w_{2}$\cite{zudov4}. The unexpected result
consists in a $\rho_{xx}$ response which is clearly modulated in the
oscillations amplitude. This modulation results to be tunable by
increasing intensity of one of the MW sources keeping the other
constant. All these experimental evidences establish real challenges
for the theoretical models presented to date
\cite{girvin,dietel,lei,ryzhii,rivera,shi,andreev,ina,mirlin}.
Considering that all these models are not able to achieve consensus
about the true origin of these striking phenomena, the new
experimental results can be regarded as crucial tests for theories,
for the existing ones, and for the ones to come. However some
theoretical contributions are already being presented giving
explanation for some of the experimental outcomes. We can stress
theoretical proposals for temperature and high MW intensity
dependence\cite{ina2,lei2}, absolute negative
conductivity\cite{ina3,ahn}, $\rho_{xx}$ immunity to the sense of
circular polarization of MW radiation \cite{dietel2,gumbs} and
finally one very recent proposal regarding $\rho_{xx}$ response to
bichromatic MW radiation\cite{lei3}

In this letter we report a theoretical explanation to the
$\rho_{xx}$ modulated amplitude of a 2DES when is illuminated for
two different MW radiations with different frequencies $w_{1}$ and
$w_{2}$. In a recently presented model by the authors\cite{ina},
it was demonstrated that a 2DES subjected to a perpendicular
magnetic field and MW radiation, Larmor orbit centers oscillate
with the same frequency as the MW field: {\it MW driven Larmor
orbits}. If we now apply two MW fields to the sample the
consequence is that electronic orbit centers are subjected
simultaneously to two oscillatory movements with the frequencies
of the MW fields $w_{1}$ and $w_{2}$. The outcome is the
superposition of both oscillations giving rise to a oscillatory
movement in the orbits center whose amplitude is modulated in the
way of pulses. Pulses have the well-known frequency of
$\frac{w_{1}-w_{2}}{2}$ whereas the new oscillatory movement goes
with $\frac{w_{1}+w_{2}}{2}$.

Following the {\it MW driven Larmor orbits model}, we first obtain
the exact expression of the electronic wave vector for a 2DES in a
perpendicular magnetic field $B$ and two MW
sources\cite{ina,kerner,park}:
\begin{eqnarray}
&&\Psi(x,t)=\phi_{n}(x-X-x_{cl}(t),t) \times  exp
\left[i\frac{m^{*}}{\hbar}\frac{dx_{cl}(t)}{dt}[x-x_{cl}(t)]+
\frac{i}{\hbar}\int_{0}^{t} {\it L} dt'\right]\nonumber  \\
&&[ \sum_{m=-\infty}^{\infty} J_{m}\left[ \frac{eE_{1}X }{\hbar
w_{1}} \right]
\left(\frac{1}{w_{1}}+\frac{w_{1}}{\sqrt{(w_{c}^{2}-w_{1}^{2})^{2}+\gamma^{4}}}\right)
e^{ipw_{1}t} + \nonumber\\
&&\sum_{n=-\infty}^{\infty} J_{n}\left[ \frac{eE_{2}X }{\hbar
w_{2}} \right]
\left(\frac{1}{w_{2}}+\frac{w_{2}}{\sqrt{(w_{c}^{2}-w_{2}^{2})^{2}+\gamma^{4}}}\right)
e^{ipw_{2}t}]
\end{eqnarray}
where $\gamma$ is a phenomenologically-introduced damping factor
for the electronic interaction with acoustic phonons\cite{ina2},
$e$ is the electron charge, $\phi_{n}$ is the solution for the
Schr\"{o}dinger equation of the unforced quantum harmonic
oscillator, $w_{1}$ and $w_{2 }$ are the MW frequencies, $E_{1}$
and $E_{2}$ are the intensities  for the MW fields, $w_{c}$ is the
cyclotron frequency, $X$ is the center of the orbit for the
electron motion and  $x_{cl}$ is the classical solution of a
forced harmonic oscillator driven by two different time dependent
forces:\\
$x_{cl}(t)=x_{1}(t)+x_{2}(t)= \\
=\frac{e
E_{1}}{m^{*}\sqrt{(w_{c}^{2}-w_{1}^{2})^{2}+\gamma^{4}}}\cos w_{1}t+
\frac{e E_{2}}{m^{*}\sqrt{(w_{c}^{2}-w_{2}^{2})^{2}+\gamma^{4}}}\cos
w_{2}t =\\
=A_{1}\cos w_{1}t +  A_{2}\cos w_{2}t $

$L$ is the classical Lagrangian and $J_{m}$ and $J_{n}$ are Bessel
functions. If $w_{1}$ is not very different from $w_{2}$ and the
MW fields intensities are equal, then we can write,\\ $A_{1}\simeq
A_{2}=A$ and therefore:

$x_{cl}(t)=A [\cos w_{1}t+ \cos w_{2}t]= 2A \cos\left
[\frac{1}{2}(w_{1}-w_{2})t\right ] \cos \left
[\frac{1}{2}(w_{1}+w_{2})t\right] $
\\showing that now the oscillatory movement for the Larmor orbits
center presents modulated amplitude with a frequency given by
$\frac{1}{2}(w_{1}-w_{2})$ whereas the main oscillation goes like
$\frac{1}{2}(w_{1}+w_{2})$.

 Now we introduce the scattering
suffered by the electrons due to charged impurities randomly
distributed in the sample\cite{ridley,ina}.  Following the model
described in ref. [19], firstly we calculate the electron-charged
impurity scattering rate $1/\tau$ (being $\tau$ the scattering
time). Secondly we find the average effective distance advanced by
the electron in every scattering jump, that in the case of two MW
sources is given by:\\
 $\Delta X^{MW}=\Delta X^{0}+ A_{1}\cos
w_{1}\tau+A_{2}\cos w_{2}\tau$, where $\Delta X^{0}$ is the
effective distance advanced when there is no MW field present.
Again if  $A_{1}\simeq A_{2}=A$ we can write:\\ $\Delta
X^{MW}=\Delta X^{0}+2A \cos\left
[\frac{1}{2}(w_{1}-w_{2})\tau\right ] \cos \left
[\frac{1}{2}(w_{1}+w_{2})\tau\right] $.
 Finally the longitudinal
conductivity $\sigma_{xx}$ can be calculated: $\sigma_{xx}\propto
\int dE \frac{\Delta X^{MW}}{\tau}(f_{i}-f_{f})$,  being $f_{i}$ and
$f_{f}$ the corresponding distribution functions for the initial and
final Landau states respectively and $E$ energy. To obtain
$\rho_{xx}$ we use the relation
$\rho_{xx}=\frac{\sigma_{xx}}{\sigma_{xx}^{2}+\sigma_{xy}^{2}}
\simeq\frac{\sigma_{xx}}{\sigma_{xy}^{2}}$, where
$\sigma_{xy}\simeq\frac{n_{i}e}{B}$ and $\sigma_{xx}\ll\sigma_{xy}$.


In Fig. 1, we present calculated results for experimental
frequencies\cite{zudov4} ($47 GHz$ and $31 GHz$). In panel a) we
represent all the cases (bichromatic and both monochormatic)
together for comparison. In the rest of panels we represent each
calculated response separately. According to our model the
surprising profile obtained for $\rho_{xx}$ response is a reflex
of the amplitude modulated oscillatory movement of the Larmor
orbit centers, when they are under the influence of both  MW
fields. In fact in the experimental graphs \cite{zudov4} and in
the calculated Fig 1., at least one modulated pulse can be seen.
This pulse can also be observed more clearly in Fig. 2, where we
represent same situation as in Fig 1., but with different
frequencies: ($57 GHz$ and $38 GHz$). Our results are in good
agreement with experiments.

In Fig. 3, we represent in the bottom graph  calculated
bichromatic magnetoresistivity $\rho_{xx}$ as a function of $B$,
for experimental frequencies at increasing MW-intensity for the
case of $47 GHz$. In the top graph of Fig. 3, we present
mathematical functions which simulate $\rho_{xx}$ behavior for
monochromatic and bichromatic driving forces, considering one of
them with increasing magnitude. In both graphs it can be observed
that a minimum shows up when one of the driving forces is
increased. This is because increasing only one of the intensities
($47 GHz$ in this case), the bichromatic response will tend to be
similar to the monochromatic one. We have considered that,
according to our model (see ref. [19]), $\rho_{xx}$ depends on $B$
like $\rho_{xx}\propto B \cos(w \tau) = B
\cos\left(S_{c}\frac{w}{B}\right) $, where $S_{c}$ is a sample
dependent scattering term. In the case of two simultaneously  MW
driving forces with similar intensities: $\rho_{xx}\propto
B\cos\left [\frac{1}{2}\frac{(w_{1}-w_{2})}{B}S_{c}\right ] \cos
\left [\frac{1}{2}\frac{(w_{1}+w_{2})}{B}S_{c}\right] $.
Considering the last expression we have tested our current theory
using the experimental data (Fig. 2 of ref. [11]). The aim is to
eventually obtain the experimental frequencies used  ($47 GHz$ and
$31 GHz$). In order to do that first we have measured in the
experimental figure the $B^{-1}$ periodicity in monochromatic and
bichromatic graphs and also the width of the corresponding
bichromatic pulse. Once obtained this information, and using our
expressions for the $\rho_{xx}$ dependence with $B$ we have been
able to reach numerical values for the pulse and main oscillation
frequencies. For the pulse we have obtained a value of $9.3 GHZ$,
and for the main frequency $38.9 GHz$. Comparing these values with
the ones obtained directly from the experimental frequencies,
$\frac{(47-31)GHz}{2} = 8 GHz$ for the pulse and
$\frac{(47+31)GHz}{2}  = 39 GHz$ for the main frequency, we can
see that  the agreement finally achieved is quite reasonable.

In conclusion we have demonstrated that the experimental
results\cite{zudov4} regarding the $\rho_{xx}$ modulated response of
a 2DES subjected to a $B$ and bichromatic MW can be explained in
terms of the electronic orbit centers being driven by the
superposition of two harmonic oscillatory movements with the
frequencies of the microwave sources.

This work was supported by the MCYT (Spain) grant MAT2005-06444,
the ``Ramon y Cajal'' program (J.I.). and the EU Human Potential
Programme HPRN-CT-2000-00144.

\newpage
\clearpage

Figure 1 caption: Calculated magnetoresistivity $\rho_{xx}$ as a
function of $B$, for experimental MW-frequencies\cite{zudov4}: $47
GHz$ and $31 GHz$. a) Calculated MW responses for all the
frequencies considered, monochromatic and bichromatic MW sources,
all together for comparison . b) Bicromatic response ($47 GHz$ and
$31 GHz$). c) Monochromatic $47 GHz$. d)monochromatic $31 GHz$.
\newline

Figure 2 caption: Same as Fig. 1 but for MW frequencies $57 GHz$
and $38 GHz$
\newline

Figure 3 caption: a). Bottom graph: calculated magnetoresistivity
$\rho_{xx}$ as a function of $B$, for experimental frequencies at
increasing MW-intensity for the $47 GHz$ MW source. Top graph:
Mathematical functions to simulate $\rho_{xx}$ behavior for
monochromatic and bichromatic driving forces, considering one of
them with increasing magnitude.
\newline

\newpage
\begin{figure}
\centering \epsfxsize=3.5in \epsfysize=3.5in
\epsffile{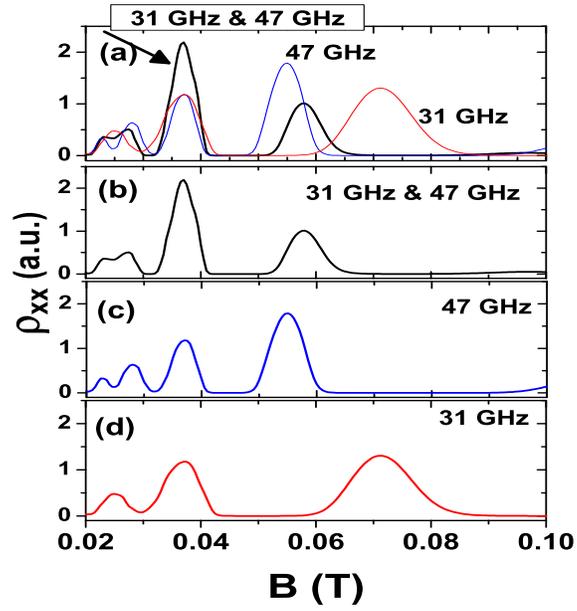}
\caption{Calculated magnetoresistivity $\rho_{xx}$ as a function
of $B$, for experimental MW-frequencies\cite{zudov4}: $47 GHz$ and
$31 GHz$. a) Calculated MW responses for all the frequencies
considered, monochromatic and bichromatic MW sources, all together
for comparison . b) Bicromatic response ($47 GHz$ and $31 GHz$).
c) Monochromatic $47 GHz$. d)monochromatic $31 GHz$.}
\end{figure}
\newpage
\begin{figure}
\centering\epsfxsize=3.5in \epsfysize=3.5in
\epsffile{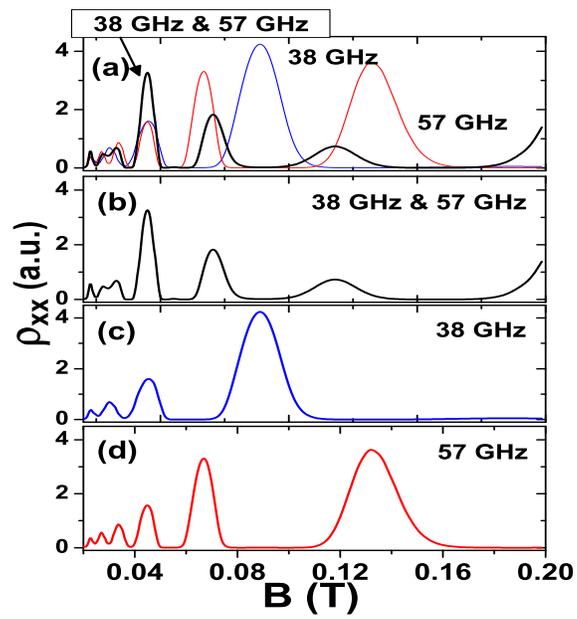}
\caption{Same as Fig. 1 but for MW frequencies $57 GHz$ and $38
GHz$.}
\end{figure}
\newpage
\begin{figure}
\centering\epsfxsize=3.5in \epsfysize=4.0in
\epsffile{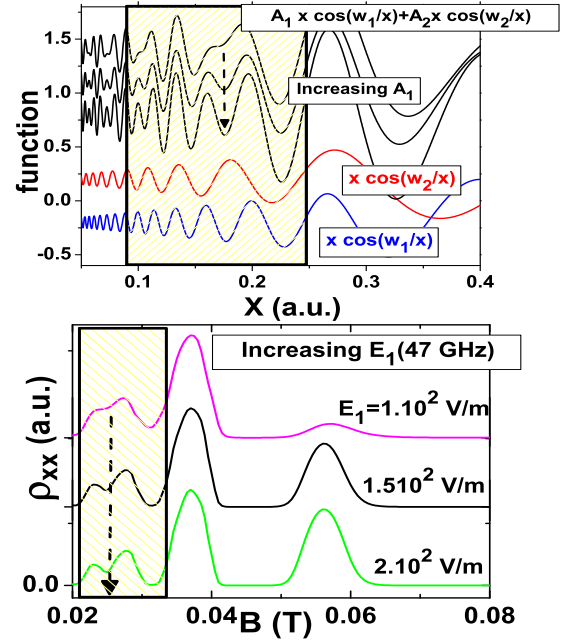}
\caption{a). Bottom graph: calculated magnetoresistivity
$\rho_{xx}$ as a function of $B$, for experimental frequencies at
increasing MW-intensity for the $47 GHz$ MW source. Top graph:
Mathematical functions to simulate $\rho_{xx}$ behavior for
monochromatic and bichromatic driving forces, considering one of
them with increasing magnitude.}
\end{figure}
\end{document}